\documentstyle{aa}               
\input epsfig.sty
\input epsf.sty

\newcommand{\cao}{\c{c}\~{a}o}
\newcommand{\ece}{\^{e}}

\newcommand{\om}{\delta}

\begin{document}
 

    \title{An analytical MHD wind model with latitudinal dependences 
           obtained using separation of the variables}
  
   \author{J.J.G. Lima
           \inst{1,2}
   \and    E.R. Priest
           \inst{3}
   \and    K. Tsinganos
           \inst{4}
          }
 
   \offprints{J.J.G. Lima}
 
   \institute{Centro de Astrof\'{\i}sica,
              Universidade do Porto,
              Rua das Estrelas,
              4150 -- 762 Porto, Portugal\\
              \email{jlima@astro.up.pt}
   \and       Departamento de Matem\'{a}tica Aplicada, Faculdade de Ci\^{e}ncias,
              Universidade do Porto, Portugal
   \and       School of Mathematical and Computational Sciences,
              University of St.Andrews,
              St.Andrews KY16 9SS,
              Scotland, UK
              \email{eric@mcs.st-and.ac.uk}
   \and       Department of Physics, University of Crete and FORTH,   
              P.O. Box 2208, 710 03 Heraklion, Crete, Greece\\
              \email{tsingan@physics.uoc.gr}
             }
 
   \date{Received February 20, 2001; accepted March 12, 2001}
 
   \abstract{A new class of analytical 2-D solutions of the full set of the steady 
   magnetohydrodynamic (MHD) equations, describing an axisymmetric helicoidal 
   magnetized outflow originating from a rotating central object, is presented.
   The solutions are systematically obtained via a nonlinear separation of the 
   variables in the momentum equation.
   The analysis yields three parameters which measure the anisotropy in the 
   latitudinal distribution of various flow quantities.  
   Topologically, the wind speed is controlled by an X-type critical point that 
   acts to filter out a single wind-type branch and the Alfv\'en singularity. 
   The solutions can be regarded as an extension outside the equatorial plane 
   of the Weber \& Davis (1967) model of magnetized winds but with a variable 
   polytropic index. 
}

\titlerunning{An analytical MHD wind model with latitudinal dependences}
\authorrunning{J.J.G. Lima et al.}
\maketitle
\section{Introduction}
\label{s1}
 
Recent detailed observations of the solar wind (e.g., via Ulysses and SoHO, 
Feldman et al. 1996) and other astrophysical outflows (e.g., via the 
Hubble Space Telescope, Biretta 1996) have highlighted the need to provide a 
satisfactory description of cosmic MHD outflows which deals adequately and 
simultaneously with their: 
(i) spatial 2-D (or 3-D) character, (ii) nonlinear dynamics, 
(iii) detailed energy balance and (iv) time-dependent nature.  
At the moment this task is far from being completed and still remains a 
challenge for the future. The present treatment is not going to change this 
fact that we do not have available to this day a satisfactory MHD solar/stellar 
wind description. Hence, by necessity, in the present paper we shall constrain 
our efforts only to the above items (i)-(ii). 
        
The first models of astrophysical outflows were {\it polytropic} and  
1-D (Parker 1958, 1963), or, all forces perpendicular to the equatorial plane were 
ignored (Weber \& Davis 1967).
The discovery of flows through coronal holes called for the investigation 
of quasi-two dimensional polytropic models at first (Kopp \& Holzer 1976, 
Habbal \& Tsinganos 1983) and fully 2-D numerical models later (Pneuman \& Kopp 1971, 
Steinolfson et al. 1982,  Nerney \& Suess 1975, Sakurai 1985). 
Recently, time-dependent studies of thermally and/or magneto-centrifugally-driven 
polytropic wind-type outflows provided new and 
promising results in relation to the question of the degree of 
collimation of the outflow (Ouyed \& Pudritz 1997, Keppens \& Goedbloed 1999, 
Ustyugova et al. 1999, Tsinganos \& Bogovalov 2000, Krasnopolsky et al. 2000).
{\it Non-polytropic} wind modelling on the other hand, with energy and momentum addition 
and finite thermal conductivity has been increasingly used because observations 
have highlighted the fact that the acceleration of the solar wind in high-speed 
streams (Feldman et al. 1996, Giordano et al. 2000) does imply energy and momentum 
addition in the solar corona (Leer \& Holzer 1980,  Steinolfson 1988, 
Suess et al, 1996, Hansteen et al. 1997,  Wang et al. 1998).  

A fully 2-D MHD modelling of steady hydromagnetic wind-type outflows 
in open magnetic fields has been introduced (Low \& Tsinganos 1986, Tsinganos \& Low 1989) 
with a variable polytropic index $\gamma$ wherein it was shown that the density needs to 
decrease with latitude for an accelerated wind. This is because a dipolar
magnetic field needs to be kept open by a pressure that must decrease towards
the pole. If the density does not vary with latitude, there is a smaller
pressure gradient to drive the flow near the pole, exactly where the magnetic 
field is open
to allow the wind to escape. The resulting acceleration is too low,
since gravity dominates, and the flow does not reach a high enough terminal
speed. The only way out is to allow the density to increase with latitude,
faster than the pressure does (Hu \& Low 1989).
Subsequently, solutions of the MHD equations with a latitudinally-dependent 
density and a helicoidal geometry of the streamlines were analysed   
(Tsinganos \& Trussoni 1991, hereafter referred to as TT91), but    
the latitudinal dependence of the different quantities was assumed {\it a priori}.
Under the same assumptions, this work was later continued (Trussoni \& Tsinganos 1993,
Trussoni et al. 1997). Other analytical solutions which have used similar assumptions
have concentrated on the asymptotic analysis and its links to the intitial boundary conditions 
and were able to derive a criterion for the collimation of winds into jets 
(Sauty \& Tsinganos 1994, Sauty et al. 1999).

In this paper we shall deduce from the governing MHD equations via a nonlinear 
separation of the variables the latitudinal variation of the density and other  
relevant physical quantities of the wind, instead of adopting {\it a priori} 
for them a specific form, as it was done in TT91 and in subsequent papers of that series.  
This shall also further extend previous work (Lima \& Priest 1993, hereafter 
referred as Paper I), which deduced a general class of solutions of the 
hydrodynamic equations relevant to stellar winds with a helicoidal geometry.  
Sect. 2 is devoted to the method of solution, Sect. 3 deals 
with its parametric study and finally Sect. 4 ends  with a discussion of the results.

\section{Method of solution}
\subsection{Basic equations and assumptions}
\label{s2.1}

We shall seek 2-D solutions of the set of steady MHD equations describing the
dynamical interaction of an inviscid, compressible and highly conducting 
plasma with an axially symmetric magnetic field created by a central 
rotating object. These include Maxwell's equation for the divergence of the
magnetic field, the induction equation in the limit of large conductivity
and the equations for conservation of mass and momentum (Priest 1982).
The pressure is related to the density and temperature through the classical
ideal gas law $p=(2k_B/m)\rho T$ while the temperature is obtained from an energy 
conservation equation such as the first law of thermodynamics, including a 
heating/cooling term. 

The system of MHD equations can be closed if the heating function 
is functionally related to the thermodynamic variables and the flow field.  
One such simple example which has been widely used in the literature is when we have
a polytropic law between pressure and density. 
The alternative approach that we follow here is to find some 
other functional relationship 
between  the heating rate, pressure, density and flow speed such that the variables 
are separable. 

The obvious choice of coordinate system appropriate to this type of 
problem is that of spherical polar coordinates $(r,\theta,\phi)$, with 
$\theta$ as the co-latitude. 
With axisymmetry, in order to make the above equations more tractable from an analytical point of 
view, the simplifying assumption will be made that the meridional components
of both the velocity and magnetic field can be neglected 
(i.e. $V_{\theta}=B_{\theta}=0$). Under this assumption the projection of
the field lines on the poloidal plane corresponds to radial lines. Note that it 
has been
recently shown that this assumption of radiality is a good one, at least in
the case of the solar wind (Wang et al 1998, Tsinganos \& Bogovalov 2000).

Then, the system of MHD equations can be re-written explicitly in the following way:
\begin{equation}\label{11}
{\partial\over\partial r}(B_r r^2)=0,
\end{equation}

\begin{equation}\label{12}
{\partial\over\partial r}(rV_{\phi}B_r-rV_rB_{\phi})=0,
\end{equation}

\begin{equation}\label{13}
{\partial\over\partial r}(\rho r^2 V_r)=0,
\end{equation}

\begin{equation}\label{14}
\rho V_r{\partial V_r\over\partial r}-\rho{V_{\phi}^2\over r}=
-{\partial p\over \partial r}
-{B_{\phi}^2\over 4\pi r}
-{B_{\phi}\over 4\pi}{\partial B_{\phi} \over\partial r}
-{\rho GM\over r^2},
\end{equation}

\begin{equation}\label{15}
\rho V_{\phi}^2{\cos\theta\over \sin\theta}=
{\partial p\over\partial\theta}
+{B_r\over 4\pi}{\partial B_r\over\partial\theta}
+{B_{\phi}^2\over 4\pi}{\cos\theta\over \sin\theta}
+{B_{\phi}\over 4\pi}{\partial B_{\phi}\over\partial\theta},
\end{equation}

\begin{equation}\label{16}
\rho V_r{\partial V_{\phi}\over\partial r}
+\rho{V_r V_{\phi}\over r}=
{B_r B_{\phi}\over 4\pi r}
+{B_r\over 4\pi}{\partial B_{\phi} \over\partial r}.
\end{equation}

\subsection{A technique based on a separation of the variables}
\label{s2.2}

At this stage, we introduce a second simplifying assumption, namely that
the variables are separable in $r$ and $\theta$. This will transform the 
above system of partial differential
equations into a system of ordinary differential equations which are 
analytically more tractable. 

Denoting by $r_0$ the radius at the base of the atmosphere, we can 
non-dimensionalise all quantities with respect to their values at this 
reference level. In particular we set $R=r/r_0$. Using the assumption of 
separation of variables, we can write the radial velocity, the azimuthal 
velocity and the azimuthal magnetic field as, respectively,
\begin{equation}\label{18.z}
V_r(R,\theta)=V_0 Y(R)v_r(\theta),
\end{equation}
\begin{equation}\label{19a}
V_{\phi}(R,\theta)=V_1 A(R)v_{\phi}(\theta),
\end{equation}
\begin{equation}\label{19b}
B_{\phi}(R,\theta)=B_1 M(R)b_{\phi}(\theta),
\end{equation}
where $V_0$, $V_1$ and $B_1$ correspond to their reference values. Note
that, at this stage, the functions 
$v_r(\theta)$, $v_{\phi}(\theta)$ and $b_{\phi}(\theta)$ are 
completely arbitrary. 
>From Eqs.(\ref{11}), (\ref{13}) we must also have
\begin{equation}\label{20.z}
B_r(R,\theta)=B_0 {b_r(\theta)\over R^2},\;\;\;\;\;\;\;\;
\rho(R,\theta)=\rho_0 {g(\theta)\over Y R^2},
\end{equation}
with $B_0$ and $\rho_0$ as the values of the radial magnetic field and 
density at the reference level, while $b_r(\theta)$ and $g(\theta)$ are 
as yet to be determined. Note that the radial magnetic field has a monopole 
geometry modified by the presence of $b_r(\theta)$.
 
Using the same technique described in Paper I,
let us eliminate the pressure term between the $r$- and $\theta$-
components of the momentum equation, Eqs.(\ref{14}), (\ref{15}),
by differentiating the first
one with respect to $\theta$ and the second one with respect to $r$, and 
then adding. The resulting expression is
\begin{eqnarray}
{\partial\over\partial\theta}\left(\rho{V_{\phi}^2\over r}\right)-
{\partial\over\partial\theta}\left(\rho V_r{\partial V_r\over\partial r}\right)-
{1\over 4\pi}{\partial\over\partial\theta}\left(
{B_{\phi}^2\over r}\right) & \nonumber\\
\nonumber\\
-{1\over 4\pi}{\partial\over\partial\theta}\left(
B_{\phi}{\partial B_{\phi} \over\partial r}\right)-
{\partial\over\partial\theta}\left({\rho GM\over r^2}\right) & \nonumber\\ 
\label{18} \\
-{\partial\over\partial r}\left(\rho V_{\phi}^2{\cos\theta\over \sin\theta}\right)+
{1\over 4\pi}{\partial\over\partial r}\left(
B_r{\partial B_r\over\partial\theta}\right)& \nonumber\\
\nonumber\\
+{1\over 4\pi}{\partial\over\partial r}\left(
B_{\phi}^2{\cos\theta\over \sin\theta}\right)+
{1\over 4\pi}{\partial\over\partial r}\left(
B_{\phi}{\partial B_{\phi}\over\partial\theta}\right) & =0. \nonumber
\end{eqnarray}
\\
Under the assumption of separation of variables, the above equation will be
transformed into an ordinary differential equation involving functions of $R$
alone. For that purpose, the functions of $\theta$ in each term will be set 
proportional to each other. We should keep in mind, however, that this last 
procedure may not give us the most general separable solution in the sense 
that some terms could be also set proportional to the sum of the others. 
Thus, from the last two terms, we may write
\begin{equation}
b_{\phi}{{\rm d} b_{\phi}\over {\rm d}\theta}=\epsilon 
b_{\phi}^2{\cos\theta\over \sin\theta},
\end{equation}
which implies that
\begin{equation}\label{18.1}
{{\rm d}\over {\rm d}\theta}\left({b_{\phi}^2\over\sin^{2\epsilon}\theta}\right)=0,
\end{equation}
and so
\begin{equation}\label{18a}
b_{\phi}(\theta)=\sin^{\epsilon}\theta,
\end{equation}
where $\epsilon$ is an arbitrary constant. 
Note that the constant of integration has been set to unity,
without loss of generality. Any other value for this constant can be incorporated 
into the radial dependence of  $B_{\phi}(R,\theta)$. In what follows, the choice of
the constants of proportionality between different terms as well as the 
constants of integration will be made so as to obtain the simplest possible
solutions, without loss of generality.
Comparing the $7^{th}$ and $8^{th}$ terms, we can put
\begin{equation}\label{18aa}
b_{r}{{\rm d}b_{r}\over {\rm d}\theta}=
\mu\epsilon b_{\phi}^2{\cos\theta\over \sin\theta},
\end{equation}
which implies
\begin{equation}\label{18b}
b_r(\theta)=\sqrt{1+\mu \sin^{2\epsilon}\theta},
\end{equation}
where $\mu$ is a second  arbitrary constant.

From the $5^{th}$ and $8^{th}$ terms in Eq.(\ref{18}) we may also obtain
\begin{equation}
{{\rm d}g\over {\rm d}\theta}=2\delta\epsilon b_{\phi}^2
{\cos\theta\over \sin\theta},
\end{equation}
thus resulting in
\begin{equation}\label{18c}
g(\theta)=1+\delta \sin^{2\epsilon}\theta,
\end{equation}
in which $\delta$ is the third arbitrary constant.
As for the $6^{th}$ and $8^{th}$ terms we can write
\begin{equation}
g v_{\phi}^2=b_{\phi}^2,
\end{equation}
or, equivalently,
\begin{equation}\label{18d}
v_{\phi}(\theta)={\sin^{\epsilon}\theta\over \sqrt{1+\delta \sin^{2\epsilon}\theta}}.
\end{equation}
Finally, from the $\phi$-component of the momentum equation,
Eq.(\ref{16}), for the separation in $R$ and $\theta$ to work
we must have
\begin{equation}
gv_rv_{\phi}=b_rb_{\phi},
\end{equation}
giving
\begin{equation}\label{18e}
v_r(\theta)=\sqrt{{1+\mu \sin^{2\epsilon}\theta\over 
1+\delta \sin^{2\epsilon}\theta}}.
\end{equation}
\begin{figure}[h]
\epsfysize=5truecm
\centerline{\epsfbox[150, 550, 400, 720]{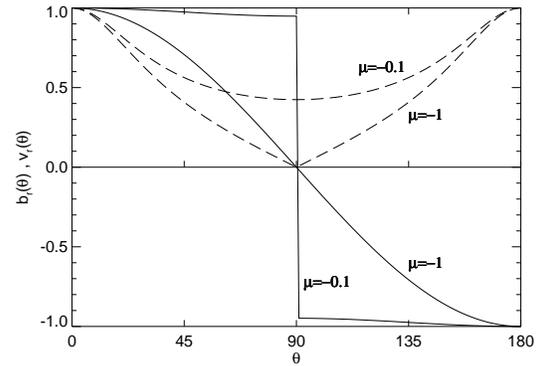}}
\caption[]{\label{f3.1}
Co-latitudinal dependence of the radial magnetic field
$b_r(\theta)$ (solid line) and of the radial velocity (dashed line)
for $\epsilon =1$ and $\delta =4$.}
\end{figure}

Note that from the four remaining terms in Eq.(\ref{18}) --- $1^{st}$, $2^{nd}$
$3^{rd}$ and
$4^{th}$ --- we obtain identities, provided the proportionality
constants are chosen accordingly.

We have thus deduced general $\theta$-dependences for the density and
hydromagnetic field, under the assumption of separation of variables,
Eqs.(\ref{18a}), (\ref{18b}), (\ref{18c}), (\ref{18d}) and (\ref{18e}), 
in terms of three arbitrary constants: ($\delta, \epsilon, \mu$). 
These parameters effectively control the anisotropy in 
the outflow, in the magnetic field and in the density distribution. 

{\it First}, the parameter $\delta$  
is related to the ratio of the density at the equator 
($\theta = \pi/2$)
to that at the pole ($\theta = 0$) and so the higher it is, the more the 
density distribution deviates from the spherically symmetric case ($\delta =0$).

{\it Second}, the parameter $\mu$ is related to the ratio of the radial 
kinetic energy density at the equator to that at the pole.  
{\it Finally}, the parameter $\epsilon$ controls the width of the profile of 
the speed and density for some fixed variation between pole and equator. 

Note that in the TT91 model, the simplest possible forms of 
$b_r(\theta)$, $g(\theta)$ and $v_r(\theta)$ were chosen 
{\it a priori} such that they were able to simulate 
existing observations of the solar wind. Their expressions 
constitute a special case of the forms deduced in this work, 
corresponding to $\epsilon=1$ and $\mu=-1$.

\subsection{Angular momentum and angular velocity}
\label{s3.3b}

Returning to the $\phi$-component of the momentum equation, Eq.(\ref{16}),
it can be manipulated to give
\begin{equation}
{\partial\over\partial r}\left(r\sin\theta V_{\phi}-r\sin\theta
{B_rB_{\phi}\over 4\pi\rho V_r}\right)=0.
\end{equation}
The solution of this equation introduces a function of $\theta$ of the form
\begin{equation}\label{20}
L(\theta)=r\sin\theta V_{\phi}-r\sin\theta {B_rB_{\phi}\over
4\pi\rho V_r},
\end{equation}
which is the total angular momentum per unit mass loss carried away by
the wind, along each flow line $\theta={\rm const}$ (see Weber \& Davis 1967).
The first term on the right-hand side of Eq.(\ref{20}) is the angular
momentum carried by the advection of the flow while the second term
represents the torque associated with the magnetic stresses.
Analogously, the solution to the induction equation, Eq.(\ref{12}), introduces
another function of $\theta$ of the form
\begin{equation}\label{21}
\Omega(\theta)={1\over r\sin\theta}\left(V_{\phi}-B_{\phi}{V_r \over B_r}\right),
\end{equation}
which corresponds to the angular velocity of the roots of the field lines
on the surface of the central object.

The azimuthal components of the velocity and magnetic fields can now be
obtained via these two functions $L(\theta)$ and $\Omega(\theta)$. 
After some straightforward manipulation we arrive at:
\begin{equation}\label{25}
V_{\phi}={r_0\over Y_*}\left({Y_*-Y\over 1-M_A^2}\right)
R \sin\theta\,\Omega(\theta) ,
\end{equation}

\begin{equation}\label{24}
B_{\phi}={B_0r_0\over V_0Y_*}
\left({R^2/R_*^2-1\over 1-M_A^2}\right)
{\Omega(\theta)\sin\theta
\sqrt{1+\delta \sin^{2\epsilon}\theta}\over R},
\end{equation}
where $M_A^2=(V_r/V_A)^2$ is the radial Alfv\'{e}n Mach number
which corresponds to the ratio of the radial velocity, $V_r$, to the radial 
Alfv\'{e}nic velocity, $V_A=B_r/\sqrt{4\pi\rho}$, and $(R_*,Y_*)$ is the
point for which $M_A=1$. At this point, $Y_*R_*^2=(M_A^0)^{-2}$, where 
\begin{equation}\label{24a}
M_A^0= {V_0\over V_0^A}
\end{equation}
is the ratio of the radial velocity to the Alfv\'{e}nic radial velocity at the base of the wind.
A physical interpretation of the radial Alfv\'{e}n Mach number $M_A$ can be found 
if we write $M_A=(\rho  V_r^2/2)/(B_r^2/8\pi)$, which
represents the ratio of kinetic to magnetic energy density. 
If $M_A\ll 1$ magnetic energy dominates, whereas if
$M_A\gg 1$ the dominant energy is kinetic. The radius $R_*$ represents
the distance at which magnetic energy ceases to dominate over the kinetic energy.
The regularity condition at $R_*$, $M_A(R_*)=1$, constrains the two functions
$L(\theta)$ and $\Omega(\theta)$: 
$L(\theta)=\Omega(\theta)r_*^2\sin^2\theta$.

\begin{figure}
\epsfysize=15truecm
\centerline{\epsfbox[150, 180, 400, 720]{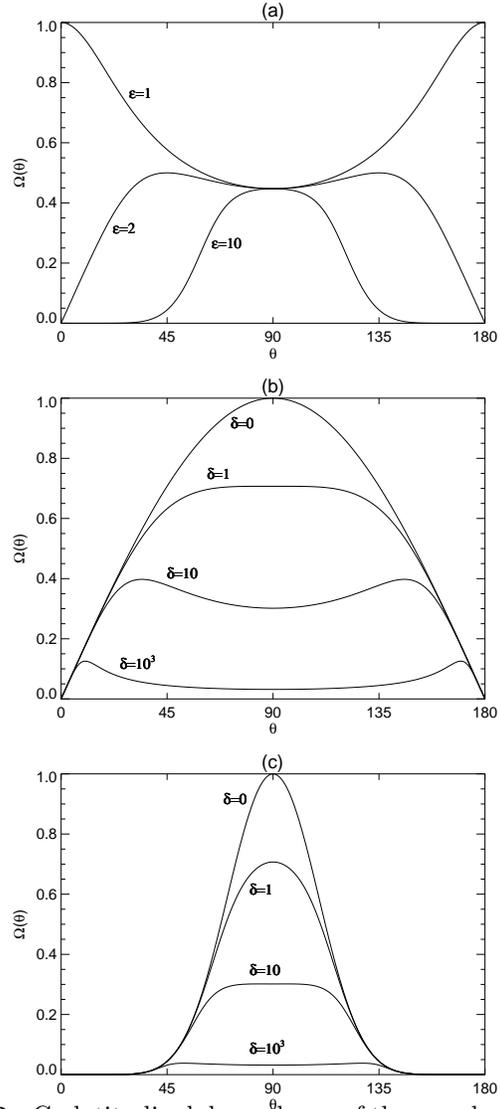}}
\vspace{-.5truecm}
\caption[]{\label{f3.3} 
Co-latitudinal dependence of the angular velocity of the roots of
the field lines $\Omega(\theta)$:
in {\bf (a)} for $\delta =4$, in {\bf (b)} for $\epsilon =2$,
in {\bf (c)} for $\epsilon =10$ }
\vspace*{-.2cm}
\end{figure}
If we now substitute Eqs.(\ref{18d}), (\ref{18a}) into Eqs.(\ref{19a}),
(\ref{19b}), respectively, and compare the resulting expressions with
 Eqs.(\ref{25}) and (\ref{24}), we can deduce that $V_0 B_1=V_1 B_0$, 
together with the general form for $\Omega(\theta)$, 
\begin{equation}\label{26}
\Omega(\theta)={\lambda V_0 Y_*\over r_0}{\sin^{\epsilon -1}\theta\over
\sqrt{1+\delta \sin^{2\epsilon}\theta}},
\end{equation}
where $\lambda=V_1/V_0=B_1/B_0$
measures the ratio of azimuthal to radial velocities at the base of the wind,
while
\begin{equation}\label{27}
L(\theta)={\lambda V_0r_0 \over (M_A^0)^2} {\sin^{\epsilon +1}\theta\over
\sqrt{1+\delta \sin^{2\epsilon}\theta}}.
\end{equation} 
The angular dependence of $\Omega (\theta)$ as the parameters $\epsilon$ 
and $\delta$ vary is worth some attention and is plotted in 
Fig. \ref{f3.3}.   
This behaviour of $\Omega (\theta  ; \epsilon , \delta )$ which emerges 
naturally from the separation of the variables in the governing equations 
can be compared, for example, with the rotation law of sunspots and solar 
photospheric magnetic fields.
Except for the regions around the poles (since from Eq. (\ref{26}), $\Omega(\theta)$ 
goes to zero for $\theta=0$ and $180^{\circ}$), the graphs of Figs. \ref{f3.3}(a,b,c) 
for $\epsilon>1$ show similarities with the angular dependence of $\Omega (\theta )$ 
arising from the yearly averaged rotation profiles of photospheric magnetic fields
(Snodgrass 1983).

\subsection{Mass and angular momentum efflux}
\label{s3.3a}

\begin{figure}[h]
\epsfysize=10truecm
\vspace*{.6cm}
\centerline{\epsfbox[140, 350, 400, 700]{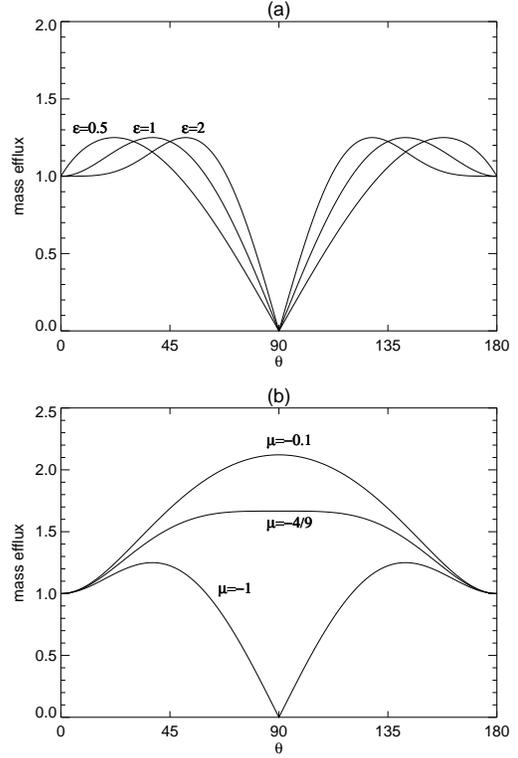}}
\caption[]{\label{f3.2}
Co-latitudinal dependence of the mass efflux $\dot{m}(\theta)$
for $\delta =4$: in {\bf (a)} for $\mu =-1$, in {\bf (b)} for $\epsilon =1$}
\end{figure}

A quantity of interest is the mass efflux ($\rho V_rr^2$) or, equivalently,
the  mass loss rate per infinitesimal solid angle $d\Sigma$ at the angle 
$\theta$ (see TT91). From the equation of conservation of mass
Eq.(\ref{13}) this has to be a function of $\theta$ alone, which we shall 
denote by $\dot{m}(\theta)$
\begin{equation}\label{18f}
\dot{m}(\theta)=\rho_0V_0r_0^2\sqrt{(1+\mu \sin^{2\epsilon}\theta )
(1+\delta \sin^{2\epsilon}\theta)}.
\end{equation}
Figure \ref{f3.2} shows the variation of the mass efflux with latitude. It vanishes
at the equator only for $\mu=-1$.
For a particular $\delta$, if
$|\mu |<\delta/(2\delta +1)$ the maximum of $\dot{m}$ occurs for
$\theta =90^{\circ}$, while if $|\mu |>\delta/(2\delta +1)$ it
occurs for $0<\theta <90^{\circ}$. This is shown in Fig. \ref{f3.2}(b)
for $\delta=4$.
We simply note in passing that the first of these angular dependences
of the mass loss is 
reminiscent of some observed intense mass loss rates that are thought to 
occur through equatorial stellar winds ({\it e.g.} from Be stars observed 
equator-on). Also, measurements of the Lyman alpha emission by the satellites 
Mariner 10, Prognoz, Ulysses and SoHO (Bertaux et al. 1997) have implied 
that there is higher mass efflux at the equator than at the pole.   

\begin{figure}[h] 
\epsfysize=10truecm
\centerline{\epsfbox[150, 350, 400, 720]{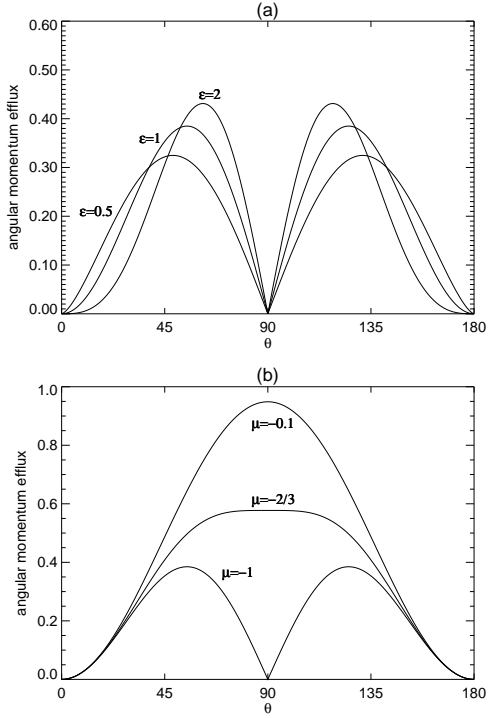}}
\vspace{-.5truecm}
\caption[]{\label{f3.4}
Co-latitudinal dependence of the angular momentum efflux
$\dot{l}(\theta)$:
in {\bf (a)} for $\mu =-1$, in {\bf (b)} for $\epsilon =1$}
\end{figure}

We may next introduce the angular momentum efflux 
$\dot{l}(\theta)\equiv\rho V_rr^2L(\theta)$ 
which is the
angular momentum loss rate per infinitesimal solid angle $d\Sigma$ at the angle
$\theta$,
\begin{equation}\label{29f}
\dot{l}(\theta)={\lambda\rho _0V_0^2r_0^3 \over (M_A^0)^2} 
\sin^{\epsilon +1}\theta\sqrt{1+\mu \sin^{2\epsilon}\theta}.
\end{equation}
The variation of $\dot{l}(\theta )$ with latitude is shown in Fig. \ref{f3.4}.
As with the mass efflux, the angular momentum efflux vanishes at the equator
for $\mu =-1$. 
In TT91 the angular momentum efflux was assumed {\it a priori},
from which the simplest possible forms for the azimuthal hydromagnetic field
were obtained. 

\subsection{Hydromagnetic field and density}
\label{s3.3d}

The azimuthal components can now be obtained 
\begin{equation}\label{28}
V_{\phi}(R,\theta)=\lambda V_0{R \sin^{\epsilon}\theta\over
\sqrt{1+\delta \sin^{2\epsilon}\theta}}
\left({Y_*-Y\over 1-M_A^2}\right),
\end{equation}

\begin{equation}\label{29}
B_{\phi}(R,\theta)=\lambda B_0{\sin^{\epsilon}\theta\over R}
\left({R^2/R_*^2-1\over 1-M_A^2}\right).
\end{equation} 
Note that $V_{\phi}$ is maximum at the equator. 
We recall that the radial hydromagnetic field and density are of the form
\begin{equation}\label{29c}
V_r(R,\theta)=V_0Y(R)
\sqrt{{1+\mu \sin^{2\epsilon}\theta\over 1+\delta \sin^{2\epsilon}\theta}},
\end{equation}

\begin{equation}\label{29d}
B_r(R,\theta)={B_0\over R^2}\sqrt{1+\mu \sin^{2\epsilon}\theta},
\end{equation}

\begin{equation}\label{29e}
\rho(R,\theta)={\rho_0\over YR^2}\left({1+\delta \sin^{2\epsilon}\theta}\right).
\end{equation}

\subsection{Balance of forces}
\label{s3.3e}

At this stage, we still have to deduce an equation for $Y(R)$, together with the
form for the pressure $p(R,\theta)$ and hence the temperature $T(R,\theta)$.
Going back to the $r$- and $\theta$-components 
of the momentum equation -- Eqs.(\ref{14}) and (\ref{15}) -- and assuming that
the variables $R$ and $\theta$ separate in these equations, we must have
\begin{equation}\label{29a}
Q(R,\theta)=Q_0(R)+Q_1(R)\sin^{2\epsilon}\theta,
\end{equation}
where $Q(R,\theta)$ is the dimensionless pressure defined by
\begin{equation}\label{29b}
p(R,\theta)={\rho_0V_0^2\over 2}Q(R,\theta).
\end{equation}
$Q_0$ represents the spherically symmetric part of the pressure, while $Q_1$ includes the
effects of the anisotropy.
Substitution of Eqs.(\ref{29a}), (\ref{29b}) into  Eqs.(\ref{14}) and
(\ref{15}) yields the following three equations for 
$Q_0(R)$, $Q_1(R)$ and $Y(R)$
\begin{eqnarray}
Q_1(R)&=&-{\mu \over (M_A^0)^2  R^4}+{\lambda^2\over\epsilon Y}
\left({Y-Y_*\over 1-M_A^2}\right)^2 \nonumber \\
& &-\left(1+{1\over\epsilon}\right)
{\lambda ^2\over  (M_A^0)^2  R^2}\left({1-R^2/R_*^2\over 1-M_A^2}\right)^2,
\label{31} \\
\nonumber \\
{{\rm d}Q_1\over {\rm d}R}&=&-{\delta\nu ^2\over YR^4}-{2\mu\over R^2}
{{\rm d}Y\over {\rm d}R}+{2\lambda ^2\over YR}\left(
{Y-Y_*\over 1-M_A^2}\right)^2 \nonumber \\
& &-{\lambda ^2 \over  (M_A^0)^2 R^2}{{\rm d}\over {\rm d}R}
\left({1-R^2/R_*^2\over 1-M_A^2}\right)^2,
\label{32} \\
\nonumber \\
{{\rm d}Q_0\over {\rm d}R}&=&-{\nu^2\over YR^4}-{2\over R^2}
{{\rm d}Y\over {\rm d}R}, \label{30}
\end{eqnarray}
in which $\nu$ is the ratio of the escape speed to the radial speed at the base of
the outflow
\begin{equation}\label{32a}
\nu={V_{esc}\over V_0}={\sqrt{2GM/r_0}\over V_0}.
\end{equation}

To understand the interplay between different forces involved
in the mechanism of this type of wind, let us describe one by one the
various terms in the above equations. All of them are written so that
the pressure gradient term is isolated on the left-hand side. 
Equation (\ref{31}) represents the force equilibrium across the
field lines.
On the right-hand side of this, the various terms represent, respectively,
the magnetic pressure, the centrifugal force and the magnetic tension.
The last two equations express the equilibrium of forces along the radial
direction. Equation (\ref{32}) shows the anisotropic terms, namely,
the anisotropic part of the gravitational and inertial forces, the centrifugal
force and the magnetic tension term, respectively. The isotropic terms are shown
in Eq.(\ref{30}). These are related to the gravitational and inertial forces, respectively.  
Equations (\ref{31}) and (\ref{32}) can be combined to give a single
expression for $Y(R)$
\begin{equation}\label{33}
{{\rm d}Y\over {\rm d}R}={F(R)\over G(R)},
\end{equation}
where
\begin{eqnarray}
F(R)&&={\delta\nu ^2\over YR^4}+{4\mu \over  (M_A^0)^2  R^5}+
{2\lambda^2\over\epsilon}{Y\over RM_A^2(1-M_A^2)^2}\times \nonumber \\
&&\left[{(1+\epsilon)M_A^2-\epsilon\over M_A^2}{R^4\over R_*^4}-
\left((2+\epsilon)M_A^2-(1+\epsilon)\right)\right]\label{34a}\\
\nonumber\\
G(R)&&=-{2\mu (M_A^0)^2 Y \over M_A^2}-{\lambda^2\over\epsilon(1-M_A^2)^2}\times \nonumber \\
&&\left[{2M_A^2-1\over M_A^4}{R^4\over R_*^4}-1\right]\label{34b}
\end{eqnarray}
The above differential equation for $Y(R)$ requires a boundary
condition. For convenience, let us choose $Y(1)=1$, which defines 
$V_0=V_r(R=1,\theta =0)$ from Eq.(\ref{29c}).
>From the classical ideal gas law we can now express temperature as
\begin{equation}\label{3.35.1}
T(R,\theta)={mV_0^2YR^2\over 2k_B(1+\delta\sin^{2\epsilon}\theta)}
Q(R,\theta).
\end{equation}

\section{Parametric study of the solution}
\subsection{Critical points}\label{s3.5.1}

In order to determine the full solution we need to solve
Eq.(\ref{33}), (\ref{34a}), (\ref{34b}) for $Y(R)$.
This is a first-order non-linear differential
equation which can be integrated numerically, using a standard routine.
A first inspection shows that this equation has a 
singular point at
$R=R_*$, $Y=Y_*$, where $M_A=1$, or in other words where the radial velocity
equals the radial Alfv\'{e}nic velocity $V_A=B_r/\sqrt{4\pi\rho}$.
The term singular point is used here in the sense that
both the numerator and denominator must vanish there.
In a more general geometry with meridional components, this point, known as
the Alfv\'{e}nic point, corresponds to the location where the poloidal
speed equals the poloidal Alfv\'{e}n speed. It is present in magnetic
wind models 
({\it c.f.} Weber \& Davis 1967, Mestel 1968)
and is a consequence of the steady-state assumption.
It delimits the magnetically dominated region
beyond which the torque exerted by the magnetic field
ceases to dominate over the angular momentum carried by the fluid.
The exact position of this point cannot be found analytically in this
case, which complicates any type of numerical treatment as we shall see later.
In a similar problem, TT91 find that expanding up to fourth
order around $(R_*,Y_*)$, all slopes are allowed.
We are thus in the presence of an improper node or star-type point.
This means that no particular solution is filtered out by the
presence of this singular point.
\begin{figure}[h] 
\epsfysize=15truecm
\hspace{.5truecm}
{\epsfig{file=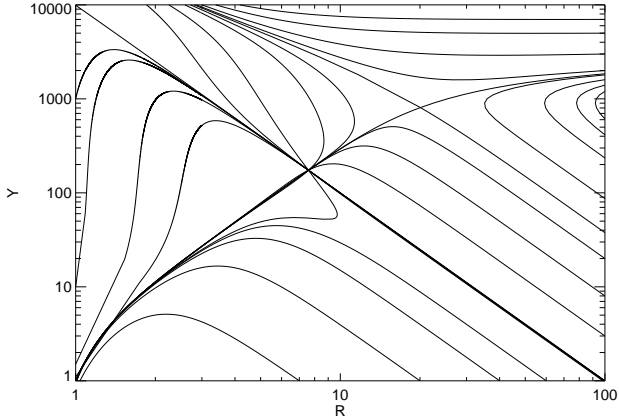, height=6truecm}}
\caption[]{\label{f3.6}
Topology of the radial dependence of the radial velocity $Y(R)$
for a solar-type highly magnetized star with $\lambda =0.5$, $\nu =120$,
$M_A^0 =0.01$, $\delta =4$, $\epsilon =1$ and $\mu =-0.1$.
Note the presence of both critical points}
\end{figure}

There is a second singular point, say at $(R_X,Y_X)$, 
found by satisfying simultaneously $F=G=0$ in Eq.(\ref{33}).
A first-order analysis around this point shows that only two slopes are allowed,
one positive and one negative, giving rise to a saddle or $X-$type point.
Thus, this point is also a critical point in the sense that it chooses
a critical solution.
In a recent work, Tsinganos et al. (1996) have shown that,  
in general, at the critical points of 
axisymmetric and self-similar solutions, the component of the flow velocity 
perpendicular to the directions of axisymmetry (the $\phi$-direction in this 
model) and self-similarity (the $\theta$-direction in this model), equals
the characteristic slow/fast MHD wave speed in that direction,
\begin{equation}\label{crit}
V_r^4 - V_r^2(V_a^2 +C_s^2) + C_s^2V^2_{a,r} = 0,
\end{equation}
with $C_s$ the sound speed and $V_{a}$, $V_{a,r}$ the total and radial 
components of the Alfv\'en speeds, respectively. 
The above equation is the equation for the speed of an MHD wave 
propagating in the r-direction. 

Thus, at the critical point where Eq. (\ref{crit}) is satisfied, the $r$-component 
of the flow speed equals to the slow or the fast MHD mode wave speed in that direction.  
Now, in low magnetisation plasmas and in the case of parallel wave propagation, the 
slow mode wave speed coincides with the Alfven speed. 
On the other hand, in high magnetization plasmas and in the case of parallel wave 
propagation,  it is the fast mode wave speed that coincides with the Alfven speed. 
In the present case we have a degeneracy in the sense that there are only two critical 
surfaces in view of the assumption that the critical surfaces are spherical
(which seems to be a good assumption, at least in the case of the solar wind -
Exarhos \& Moussas 2000). 
Note that due to the non-polytropic assumption, the sound speed is ill-defined and we 
cannot easily determine whether we are in a low- or high-magnetisation regime.  
We conjecture that here we have a situation analoguous to low magnetisation plasmas 
and at the X-type critical point the super-Alfvenic radial flow speed equals to the radial component of the 
fast MHD mode wave speed, while the slow critical transition coincides everywhere with 
the Alfv\'enic transition because of self-similarity.  

We should stress at this stage that both the Alfv\'{e}nic point $(R_*,Y_*)$
and this $X-$type point are (loosely) called equilibrium or critical points
since they satisfy the regularity condition
$F(R)=G(R)=0$. However, since the first one does not filter any solution
({\it i.e.} all slopes are allowed through it), while the second one
(as for any saddle point) selects a particular solution sometimes
referred to as the critical solution, only the X-type singularity corresponds 
to a true critical point.

We performed a numerical integration of Eqs.(\ref{33}), (\ref{34a}), 
(\ref{34b}). 
The general topology is shown in Fig. \ref{f3.6} 
for a highly magnetized medium and for 
representative values of $\mu$ and $\epsilon$.
The position of the critical points was found by an iterative
procedure. 
The integration yields a single solution crossing the $X-$type point with
positive slope and satisfying the boundary condition $Y(1)=1$.
For this critical solution, the flow starts near the star with low speeds
and connects 
to large distances where it attains large super-Alfv\'{e}nic velocities.
There is another critical solution, which
crosses the X-type point with negative slope
and is always decelerating. 
There is a limiting value of $\epsilon$
above which we couldn't find a wind solution satisfying the boundary
condition $Y(1)=1$. For example, if $\lambda =0.5$, $\nu =120$ and $\delta =4$,
there is no solution for $\epsilon\stackrel{\rm _>}{_\sim}5$, if
$M_A^0  =0.1$ and for $\epsilon\stackrel{\rm _>}{_\sim}2$, if $ M_A^0  =0.01$.
\begin{table}
\caption[]{\label{t3.1}
Location of both singular points for $\lambda =0.5$,
$\nu =120$, $\delta =4$: in {\bf (a)} for $M_A^0 =0.1$ and 
in {\bf (b)} for $ M_A^0 =0.01$}
\vspace{1cm}
\epsfysize=2.5truecm
\centerline{\epsfbox[117, 640, 500, 755]{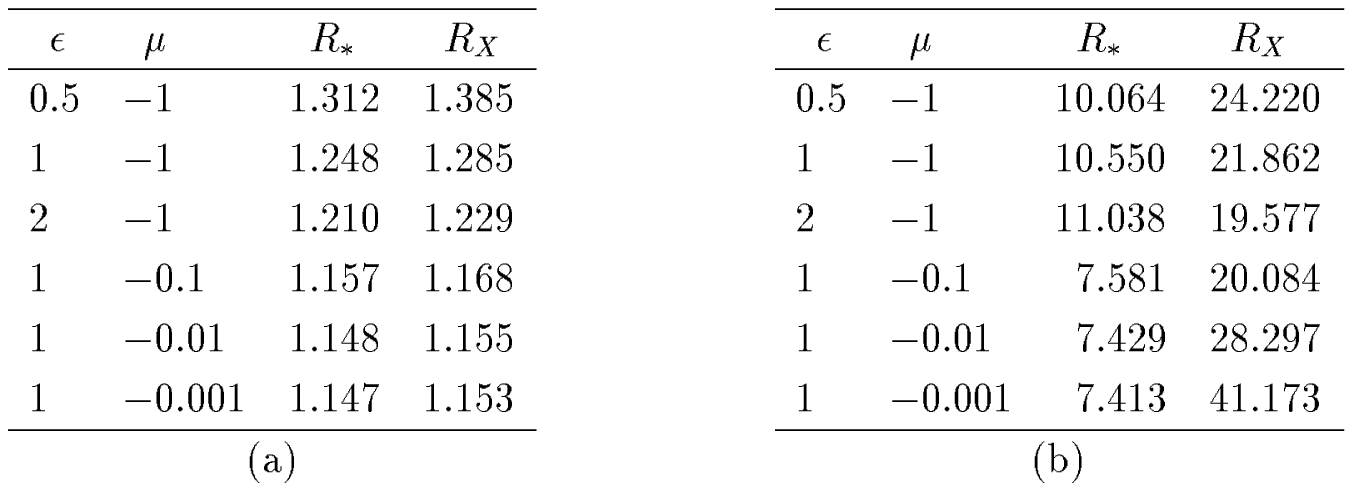}}
\end{table}

Once the positions of both $R_*$ and $R_X$ are known, all other types of
solutions can be easily found. They include breeze-type solutions that cross
the Alfv\'{e}nic point but not the X-type point and reach sub-Alfv\'{e}nic
asymptotic speeds for large distances.

Table 1 shows the positions of both singular points for $ M_A^0 =0.1$ and
$ M_A^0  =0.01$ and different values of $\epsilon$ and $\mu$.
We have taken $\lambda =0.5$, $\nu =120$ and $\delta =4$.

\subsection{Radial velocity}\label{s3.6}

\begin{figure}[h] 
\vspace*{1cm}
\epsfysize=12truecm
\hspace{-1truecm}
\centerline{\epsfbox[150, 365, 400, 680]{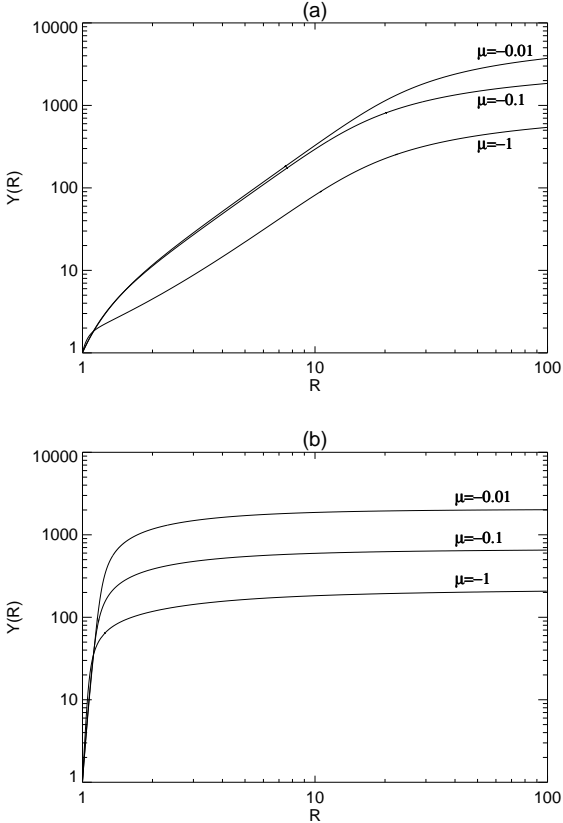}}
\vspace{-1.5truecm}
\caption[]{\label{f3.7}
Radial dependence of the dimensionless radial velocity $Y(R)$
for a solar-type star with $\lambda =0.5$, $\nu =120$, $\delta =4$, 
$\epsilon =1$: in {\bf (a)} for $ M_A^0 =0.01$, in {\bf (b)} for, $M_A^0 =0.1 $.
}
\end{figure}

The initial acceleration decreases as $M_A^0$ decreases, Figs. \ref{f3.7}(a, b), 
and also as $\lambda$ increases. In other words, the more the outflow is 
magnetised and rotating, the smaller is the initial acceleration and  
both the magnetic field and rotation are inhibiters of the initial 
acceleration of the outflow. This may be understood using a simple argument, as follows.    
Assume without loss of generality that $\delta$ = 0, such that the weight 
of a parcel of plasma is the same at any angle $\theta$, at some fixed radial 
distance $R$. Also for simplicity assume that $\mu = -1$ and $\epsilon =1$. 
Then, because of the sinusoidal dependence of the radial flow
speed with the latitude, at the equator we have static conditions such
that the weight of the plasma is balanced there mainly by the
centrifugal force and the radial component of the Lorenz force 
(the centrifugal force is always positive, i.e., outwards,
while in most of the cases examined the radial component of the magnetic
force is also positive). Then, moving at the pole and at the same radial
distance $R$, we find the same inwards gravitational plasma weight (since $\delta=0$), 
but no centrifugal and radial magnetic forces acting there due to their
${\hbox {sin}}^2 \theta$ dependence. The net result is a decelerating
force, relatively to the cases $M_A^0$ = $\lambda$ = 0.
For a mildly magnetized case ($ M_A^0  =0.1$), Fig. \ref{f3.7}(b) shows that
the initial acceleration is very high and the radial velocity rapidly
attains its asymptotic value.

The asymptotic form of $Y(R)$ for large $R$ can be obtained
after integrating Eq.(\ref{33})
\begin{equation}\label{3.52}
Y^3\simeq3\left(1+{1\over\epsilon}\right){\lambda ^2\over  (M_A^0)^6
|\mu|R_*^4}\ln R.
\end{equation}
For $ M_A^0 \ll 1$, the asymptotic speed increases logarithmically with $R$,
and $Y$ increases with a decrease of $ M_A^0 $ or a decrease
of $\epsilon$ and $|\mu|$.
When $ M_A^0 >1$, the logarithmic derivative
${\rm d}Y^3/{\rm d}\ln R$, for large $R$,
is very small (behaving like $ (M_A^0)^{-6}$),
and $Y$ can be taken as constant.
The flow is hydrodynamically dominated from $R=1$, while both singular
points, $R_*$ and $R_X$, almost coincide in the region $R<1$ ({\it i.e.},
below the base of the wind). 
We recall that in the hydrodynamic case (Paper I), $Y$ approaches
rapidly a constant value, independent of $\epsilon$.

\begin{figure}[h] 
\epsfysize=12truecm
\centerline{\epsfbox[75, 370, 300, 720]{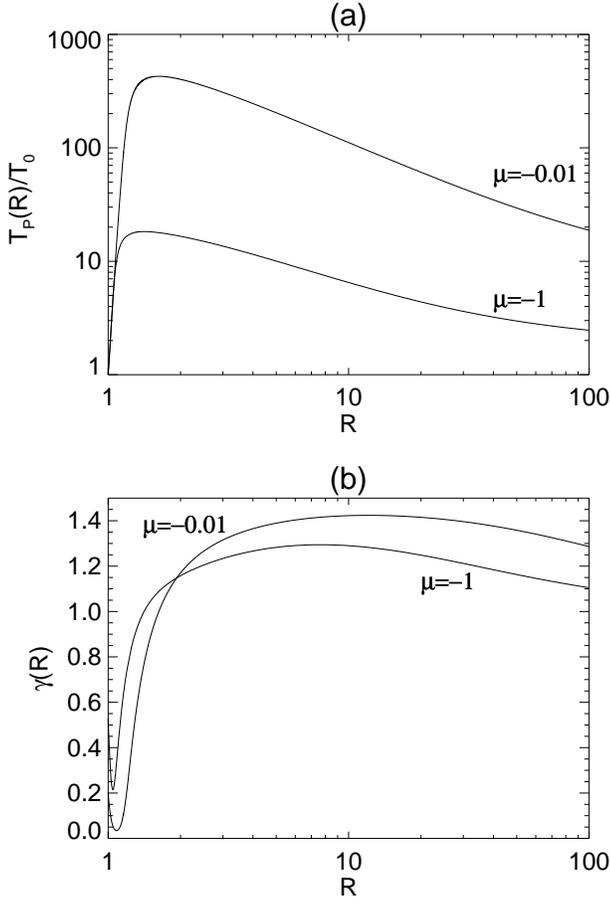}}
\caption[]{\label{f3.12}
Radial dependence of: {\bf (a)} the dimensionless temperature at the
pole $T_P(R)/T_0$, {\bf (b)} the polytropic index $\gamma(R)$,
both for $\lambda =0.5$, $\nu =120$, $ M_A^0  =0.1$, $\delta =4$, 
$\epsilon =1$ and two values of $\mu$.
}
\end{figure}

The cautious reader may have already noted that in Eq. (\ref{3.52}) 
the asymptotic radial 
speed is the product of a logarithmically diverging isothermal wind speed 
which increases like $(\ln R)^{1/3}$ (see TT91) and Michel's characteristic 
speed ($\Omega^2 F_B^2/\dot M)^{1/3}$, where $F_B$ and $\dot M$ are the magnetic 
and mass fluxes (MacGregor 1996). This is also checked by the fact that the 
effective polytropic 
index $\gamma (R \longrightarrow \infty ) \longrightarrow 1$, as seen in 
Fig. \ref{f3.12}. 

\subsection{Temperature and effective polytropic index}\label{s3.8}

In this study by not using a polytropic relationship between pressure
and density ($p\propto\rho ^\gamma$, with $\gamma$ the constant
polytropic index) we avoided constraining the exchange of energy so as
to keep $\gamma$ constant. In our approach, we can define an
effective polytropic index given by
\begin{equation}\label{56}
\gamma\equiv\left[{\partial ln p\over\partial ln \rho}\right]_{A=\mbox{const.}}
\end{equation}
for each field line $A(\theta)=\mbox{const}$. (Weber 1970, TT91).
This effective polytropic index is no longer a constant, but instead a function of $R$. 
The form of $\gamma (R)$ is closely associated with the variation of temperature
with distance from the central object. In particular, if $\gamma <1$ there
is intense heating and the temperature increases, whereas if $\gamma >1$ there
is a depletion of heating and the temperature decreases (see Fig. \ref{f3.12}).

\begin{figure}[h] 
\epsfysize=12truecm
\vspace*{-1cm}
\centerline{\epsfbox[75, 370, 300, 800]{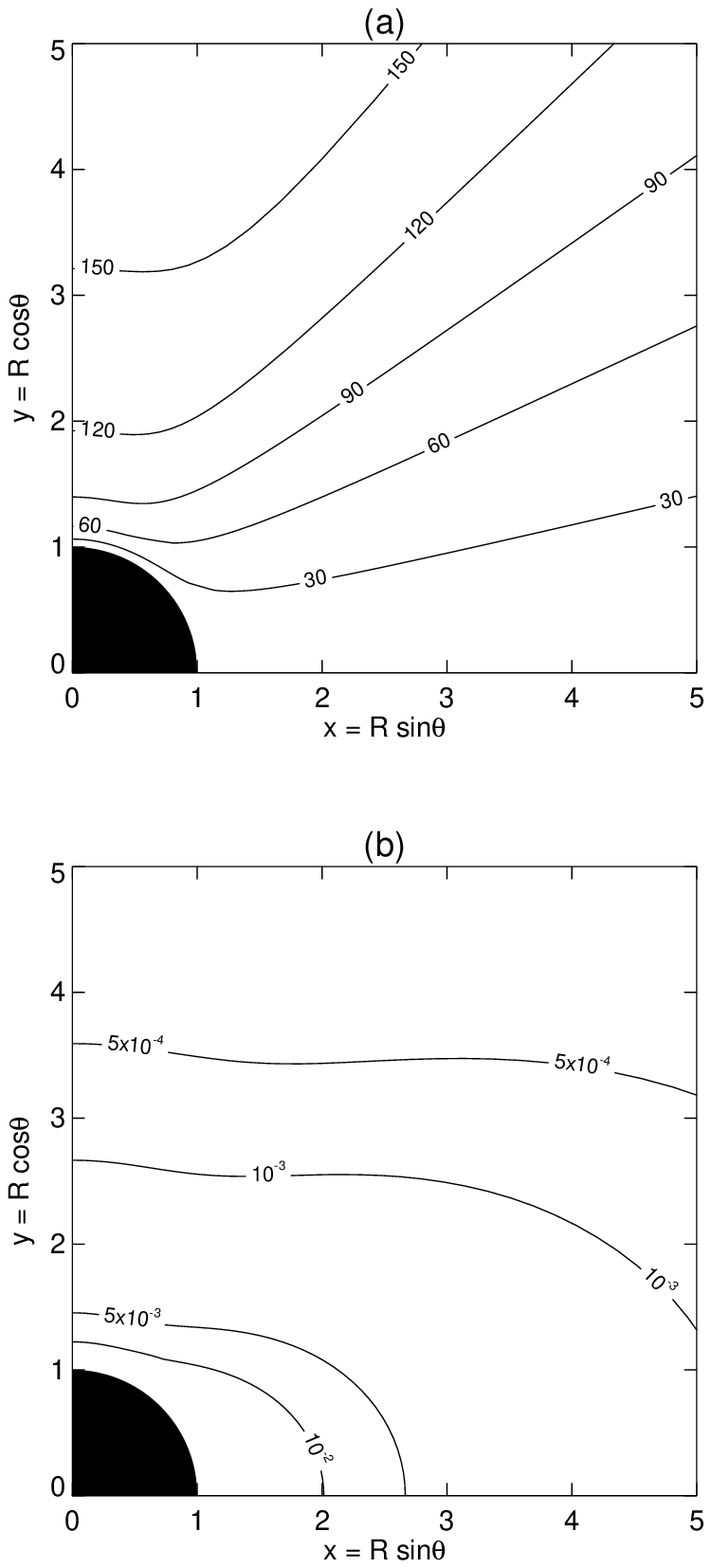}}
\vspace*{2.5cm}
\caption{\label{f3.13}
Contour plots of: {\bf (a)} the radial velocity $V_r(R,\theta)/V_0$ and  
{\bf (b)} the density $\rho(R,\theta)/\rho_0$,
for $\lambda=0.5$, $\nu=120$, $ M_A^0 =0.1$, $\om=4$, $\epsilon =2$, $\mu =-1$
}
\end{figure}

In Fig. \ref{f3.13} we have chosen to illustrate
the two-dimensional character of the solution deduced in this paper, with 
parameters corresponding to  
a mildly magnetized object ($ M_A^0 =0.1$), by plotting the 
latitudinal dependence of  the radial velocity and  density. 
The velocity, is  higher at the polar axis while the density is maximum at the equator. 
The angular dependence of these quantities is similar to the latitudinal distribution 
of the density and velocity in coronal holes.

\section{Discussion}\label{s3.10}

We have explored in this paper a new set of exact solutions of the steady
axisymmetric MHD equations relevant to stellar wind problems. 

Analogously to the case with no magnetic field (Paper I), we were able to obtain
the latitudinal dependences of the different variables
and these involve only the anisotropy parameters, $\delta$, $\epsilon$ and  
$\mu$. The parameter $\delta$ is associated with the
ratio of equatorial to polar density, $\epsilon$ controls the width of the
density and velocity profiles for a fixed variation between pole and
equator, while $\mu$ is connected to the latitudinal anisotropy 
of the distribution of  radial kinetic energy density.
The other three parameters of the problem, which come about in the solution
for the radial functions, are $\lambda$, which controls the strength of
rotation, $\nu$, a measure of the escape speed from the central object and
$ M_A^0$ which is related the strength of the magnetic pressure. 

The degree of collimation, together
with the anisotropy in the density distribution, is effectively controlled
by $\delta$, $\epsilon$ and $\mu$. These solutions could be pertinent 
to the plasma dynamics in polar coronal holes, stellar and extragalactic 
jets and even star-forming regions.
The solution with\linebreak
$\mu=-1$ has zero radial velocity at the equator
and a continuous radial magnetic field across the equator.
A discontinuity and associated current sheet arises for $\mu\neq -1$.
Also, the angular velocity of the roots of the field lines at the
surface of the central object is such that it resembles 
qualitatively solar observations only for $\epsilon >1$. 

The topology of the solutions is controlled by two distinct singular points.  
The first
one is the usual Alfv\'{e}nic point and corresponds to the location where
the radial velocity of the outflow equals the radial Alfv\'{e}nic velocity.
All solutions pass through this high-order singularity and it
corresponds to an improper node or star-type point.

There is a critical point present downstream from this Alfv\'en transition. 
It is an $X$-type critical (or saddle) point which is responsible for 
filtering a single solution corresponding to a vanishing pressure at 
infinity --- the wind-type solution. In a similar problem, Tsinganos et al. 
(1996) have shown that at this new critical point, the $r$-component of the 
flow speed equals the slow/fast MHD mode wave speeds in that direction.

Because the field/streamlines are constrained to keep 
a helicoidal geometry, the pressure gradient for large distances
has to be sufficiently large to balance the dominant tension force.
Thus, in some cases the solution is not valid outside a cone around 
the polar axis. This limitation disappears for $ M_A^0 \geq 0.05$.

There is a drastic change in the nature of the solution from high to
lower values of $ M_A^0 $.
The solutions for high $ M_A^0$ manifest all the characteristics of a
typical hydrodynamic wind, namely very large acceleration at the base
of the atmosphere 
and temperature decaying very rapidly with distance.
The magnetically dominated cases ($ M_A^0 \ll 1$)
show lower acceleration at the base of the atmosphere 
and an almost isothermal atmosphere at larger distances.

The density anisotropy ($\delta$) greatly favours the acceleration
of the wind close to the base, while the strength
of rotation ($\lambda$) and the gravitational field ($\nu$)
slows down the initial speed.
The solution far away from the source seems to be quite insensitive to $\epsilon$.
The influence of $\mu$, on the other hand, is such that a
decrease of $|\mu|$ not only diminishes the
relative importance of the magnetic effects by decreasing the size of the 
magnetic lever arm, $R_*$, but also increases the value of the
initial acceleration as well as the asymptotic
radial velocity at large distances. The 
asymptotic plasma temperature increases for lower values of $|\mu|$.

This simple but self-consistent two-dimensional solution of the MHD equations
has nevertheless some limitations. The assumption of separation of variables
could be dropped, although 
we should not forget the nonlinear character of the governing coupled partial 
differential equations. The assumption of the neglect of the meridional 
components of the magnetic field and flow ($V_{\theta}, B_{\theta}$) which 
limits their possible applications, could also be eliminated within the framework 
of the method of separation of the variables. 
Finally, as we've already discussed, the present investigation 
does not take into account the detailed energetics in the outflow.   
Nevertheless, the present study indicates that inherently non-spherically 
symmetric solutions of the MHD equations which do not have the previous 
limitations may exist. This ambitious undertaking to find them remains a challenge 
for the future.

\begin{acknowledgements}
J. Lima wishes to thank C. Sauty for his help and support and also the staff 
at FORTH in Crete for their kind hospitality. 
This work was supported by Programa Praxis XXI of the Funda\cao\
para a Ci\ece ncia e a Tecnologia, by funds from the UK Science and 
Engineering Research Council, the Erasmus Programme and the EEC RT Network 
HPRN-CT-2000-00153. 
\end{acknowledgements}


\begin{thebibliography}{}

   \bibitem{} Bertaux, J. L., Quemerais, E., Lallement, T. et al. 1997, 
             in The Corona and Solar Wind near Minimum Activity  
             (ESA SP-404) 29 

   \bibitem{} Biretta, J. A. 1996, in Solar and Astrophysical MHD Flows, 
      ed. K. Tsinganos (Kluwer Academic Publishers) 357

   \bibitem{} Exarhos, G., \& Moussas, X. 2000,
       A\&A, 356, 315

   \bibitem{} Feldman, W. C., Phillips, J. L., Barraclough, B. L., Hammond,  
      C. M. 1996, in Solar and Astrophysical MHD Flows, 
      ed. K. Tsinganos (Kluwer Academic Publishers) 265
  
  \bibitem{} Giordano,  S., Antonucci, E., Noci, G., Romoli, M., Kohl, J. L. 2000,
             ApJ Letts., preprint (astro-ph/0001257)

   \bibitem{} Habbal, S. R., \& Tsinganos, K. 1983 
      J. Geophys. Res., 88, 1965 

   \bibitem{} Hansteen, V. H., Leer, E., Holzer, T. E. 1997, ApJ, 482, 498

   \bibitem{} Hu, Y. Q., \& Low, B. C. 1989,
        ApJ,  342, 1049

   \bibitem{} Keppens, R., \& Goedbloed J. P. 1999, 
       A\&A, 343, 251 

  \bibitem{} Krasnopolsky, R., Li, Z-Y., Blandford, R. 1999, ApJ, 526, 631

   \bibitem{} Kopp, R. A., \& Holzer, T. E. 1976,
      Solar Phys., 49, 43 

   \bibitem{} Leer, E., \& Holzer T. E. 1980, J. Geophys. Res., 85, 4681   

   \bibitem{} Lima, J. J. G., \& Priest, E. R. 1993, 
       A\&A, 268, 641

   \bibitem{} Low, B. C., \& Tsinganos, K. 1986,
       ApJ, 302, 163

   \bibitem{} MacGregor,  K. B. 1996, in Solar and Astrophysical MHD Flows, 
      ed. K. Tsinganos (Kluwer Academic Publishers) 301

   \bibitem{} Mestel, L. 1968, 
       MNRAS, 138, 359

    \bibitem{} Nerney, S. F., \& Suess, S. T. 1975,
        ApJ, 196, 837
 
  \bibitem{} Ouyed, R., \& Pudritz, R. E. 1997,
       ApJ, 482, 712

    \bibitem{} Parker, E. N. 1958,
      ApJ, 128, 664

   \bibitem{} Parker, E. N. 1963,
      in Interplanetary Dynamical Processes  
      (Interscience Publishers, New York) 

   \bibitem{} Pneumann, G. W., \& Kopp, R. A. 1971,
      Solar Phys., 18, 258

   \bibitem{} Priest, E. R. 1982,
       Solar Magnetohydrodynamics (D. Reidel, Holland)

   \bibitem{} Sakurai, T. 1985, 
       A\&A, 152, 121

   \bibitem{} Sauty, C., \& Tsinganos, K. 1994, A\&A, 287, 893

   \bibitem{} Sauty, C., Tsinganos, K., Trussoni, E. 1999, A\&A, 348, 327

   \bibitem{} Snodgrass, H. B., 1983,
       ApJ, 270, 288

   \bibitem{} Steinolfson, R. S, 1988, J. Geophys. Res., 93, 14261 
   
   \bibitem{} Steinolfson, R. S., Suess, S. T., Wu, S. T. 1982,  ApJ, 255, 730 

   \bibitem{} Suess, J., Wang, A. H., Wu, S. T. 1996, J. Geophys. Res., 101, 19957  
  
   \bibitem{} Trussoni, E., \& Tsinganos, K. 1993, A\&A, 269, 589

   \bibitem{} Trussoni, E., Tsinganos, K., Sauty, C. 1997, A\&A, 325, 1099

   \bibitem{} Tsinganos, K., \& Bogovalov, S. 2000, A\&A, 356, 989

   \bibitem{} Tsinganos, K., \& Low, B. C. 1989, ApJ, 342, 1028

   \bibitem{} Tsinganos, K., \& Trussoni, E. 1991, A\&A,  249, 156 (TT91)   

   \bibitem{} Tsinganos, K., Sauty, C., Surlantzis, G., Trussoni, E., 
       Contopoulos, J. 1996, in Solar and Astrophysical MHD Flows, 
      ed. K. Tsinganos  (Kluwer Academic Publishers) 427

     \bibitem{} Ustyugova, G. V., Koldoba, A. V., Romanova, M. M., Chechetkin, V. M.,
                Lovelace, R. V. E. 1999, ApJ, 516, 221

   \bibitem{} Wang,  A. H., Wu, S. T., Suess, S. T., Poletto, G. 1998,
      J. Geophys. Res., 103(A2), 1913

   \bibitem{} Weber,  E. J. 1970,
      Solar Phys., 14, 480

   \bibitem{} Weber,  E. J., \& Davis L. J. 1967,
      ApJ, 148, 217

\end{thebibliography}
\end{document}